\documentclass[final,3p,times,11pt]{elsarticle}

\usepackage[utf8]{inputenc}
\usepackage{amsmath,amssymb,amsfonts}
\usepackage{graphicx}
\usepackage[dvipsnames]{xcolor}

\usepackage[colorlinks=true,
linkcolor=blue,
urlcolor=blue,
citecolor=blue]{hyperref}

\usepackage{tikz}
\usetikzlibrary{decorations.markings,decorations.pathmorphing,arrows.meta}
\tikzset{mom/.style={line width=0.5mm}}
\tikzset{left/.style={arrows={Stealth[scale length=0.5, scale width=1.5][sep=2pt]-}}}
\tikzset{right/.style={arrows={-[sep=-2pt]Stealth[scale length=0.5, scale
      width=1.5]}}}

\def\be{\begin{equation}}
\def\ee{\end{equation}}
\def\bes{\begin{equation*}}
\def\ees{\end{equation*}}
\def\ba{\begin{align}}
\def\ea{\end{align}}
\def\bea{\begin{eqnarray}}
\def\eea{\end{eqnarray}}

\def\f{\frac}

\def\mc{\mathcal}
\def\mb{\mathbb}

\def\v[#1]{\boldsymbol{#1}}
\def\w[#1]{\widehat{#1}}
\def\vs[#1,#2]{\boldsymbol{{#1}_{#2}}}
\def\mes[#1]{d^{3}{#1}}
\def\del{\partial}
\def\<{\langle}
\def\>{\rangle}
\def\vec[#1]{\boldsymbol{#1}}
\def\vecs[#1,#2]{\boldsymbol{{#1}_{#2}}}

\newcommand{\B}[1]{{\bar{1}}}
\newcommand{\BD}[1]{{\dot \bar{1}}}
\newcommand{\half}{\frac{1}{2}}

\def\a{\alpha}
\def\b{\beta}
\def\d{\delta}
\def\D{\Delta}
\def\e{\epsilon}
\def\ve{\varepsilon}

\def\G{\Gamma}
\def\k{\kappa}
\def\l{\lambda}
\def\L{\Lambda}
\def\m{\mu}
\def\n{\nu}
\def\N{\nabla}
\def\o{\omega}
\def\O{\Omega}
\def\p{\phi}

\def\r{\rho}
\def\s{\sigma}
\def\S{\Sigma}

\def\z{\zeta}

\journal{Nuclear Physics B}

\begin{document}

\begin{frontmatter}

\title{Hydrodynamic fluctuations and long-time tails in a fluid on an anisotropic background}

\author{Ashish Shukla} 
\ead{ashukla@perimeterinstitute.ca}
\address{Perimeter Institute for Theoretical Physics, 31 Caroline Street North, Waterloo, Ontario N2L 2Y5, Canada}

\begin{abstract}
The effective low-energy late-time description of many body systems near thermal equilibrium provided by classical hydrodynamics in terms of dissipative transport phenomena receives important corrections once the effects of stochastic fluctuations are taken into account. One such physical effect is the occurrence of long-time power law tails in correlation functions of conserved currents. In the hydrodynamic regime ${\bf k} \rightarrow 0$ this amounts to non-analytic dependence of the correlation functions on the frequency $\o$. In this article, we consider a relativistic fluid with a conserved global $U(1)$ charge in the presence of a strong background magnetic field, and compute the long-time tails in correlation functions of the stress tensor. The presence of the magnetic field renders the system anisotropic.  In the absence of the magnetic field, there are three out-of-equilibrium transport parameters that arise at the first order in the hydrodynamic derivative expansion, all of which are dissipative. In the presence of a background magnetic field, there are ten independent out-of-equilibrium transport parameters at the first order, three of which are non-dissipative and the rest are dissipative.
We provide the most general linearized equations about a given state of thermal equilibrium involving the various transport parameters in the presence of a magnetic field, and use them to compute the long-time tails for the fluid.
\end{abstract}

\begin{keyword}
Hydrodynamic fluctuations \sep Charged fluids \sep Magnetic fields \sep Non-analyticities \sep Long-time tails\\
\href{https://arxiv.org/abs/2101.10000}{arXiv:2101.10000}
\end{keyword}

\end{frontmatter}

\section{Introduction}
\label{intro}
Hydrodynamics provides an effective description of low-energy late-time dynamics of systems near  thermal equilibrium. At long time scales, the evolution and relaxation of the system is governed by the dynamics of the conserved currents associated to the system, such as the energy-momentum tensor.
In the hydrodynamic regime, the conserved currents can be expressed in an expansion in number of derivatives acting on the hydrodynamic variables, with transport parameters entering as coefficients of various terms in this expansion (see \cite{Kovtun:2012rj,  Jeon:2015dfa, Romatschke:2017ejr} for pedagogical reviews). These transport parameters capture important physical properties of the system under consideration, and can either be dissipative i.e.\ lead to entropy production, or non-dissipative.

The equations of hydrodynamics are nothing but the conservation equations for the conserved currents of the system. A salient feature of the hydrodynamic equations is the fact that they are highly nonlinear in nature. The nonlinear terms give rise to interactions between the fluctuating hydrodynamic degrees of freedom. These interactions in turn have several important consequences for the dynamics of the system. One of these is the non-analytic dependence on frequency $\o$ of the two-point correlation functions of the conserved currents of the system in the long wavelength limit ${\bf k} \rightarrow 0$. In real time, this translates to power-law dependence of correlation functions on time $t$ at late times, aptly called \textit{long-time tails} \cite{POMEAU197563, Arnold:1997gh}, as opposed to an exponential fall-off one would expect if there were no interactions. In particular, the value of the power-law exponent is independent of the microscopic details of the theory, and thus important macroscopic information can be extracted from the long-time tails, even for strongly coupled systems.

One way to think about the occurrence of fluctuations is to realize that hydrodynamics is a coarse grained macroscopic description of the system. The fluctuations that occur in the hydrodynamic degrees of freedom can thus be thought of arising as a consequence of integrating out the microscopic degrees of freedom, which may themselves have an inherent thermal randomness. This approach to hydrodynamic fluctuations, called stochastic hydrodynamics, models the fluctuations to be sourced by random microscopic currents and stresses, which are Gaussian correlated \cite{Landau1957, PhysRevA.8.423, RevModPhys.49.435}. More recently, an effective field theory (EFT) approach based on the Schwinger-Keldysh closed time path for non-equilibrium systems has been developed for studying fluctuating hydrodynamics, which takes into account the effects of fluctuations beyond the assumption of Gaussianity \cite{Crossley:2015evo, Haehl:2015uoc, Jensen:2017kzi}. See \cite{Chen-Lin:2018kfl, Jain:2020hcu, lttsk} for recent studies of analyticity properties of correlation functions and long-time tails in the Schwinger-Keldysh EFT approach.

In this paper, we initiate the study of long-time tails in a relativistic fluid with a global $U(1)$ symmetry, in the presence of a strong background magnetic field. One motivation for computing these tails could be the dynamics of the quark-gluon plasma (QGP) produced in heavy ion collisions, where the magnetic fields produced in non-central collisions can reach magnitudes of the order $10^{14}$ Tesla \cite{Kharzeev:2007jp, Skokov:2009qp, Bzdak:2011yy}. Our analysis will be similar in spirit to the ones in \cite{Arnold:1997gh, Kovtun:2003vj}, which in turn assume the fluctuations to be Gaussian in nature, just as in stochastic hydrodynamics. As pointed out in \cite{Kovtun:2003vj}, in the small ${\bf k}$ regime, the distribution of fluctuations is arbitrarily close to being Gaussian, with non-Gaussian contributions providing only sub-leading corrections to the leading long-time tails. With this in mind, we compute the linearized hydrodynamic equations about a given state of global thermal equilibrium for a relativistic fluid in the presence of a background magnetic field, and use them to compute the long-time tails in stress tensor correlation functions and their dependence on the magnetic field as well as some of the first-order transport parameters.

This article is organized as follows. In section \ref{thermodynamics}, we start by reviewing the properties of a charged fluid in thermal equilibrium in the presence of a strong background magnetic field. For a parity preserving fluid in four spacetime dimensions, which we shall be focusing on, there is one transport parameter in thermal equilibrium at one-derivative order, the magneto-vortical susceptibility $M_\O$. Out of equilibrium, there are ten more \cite{Hernandez:2017mch}. These are the three non-equilibrium non-dissipative parameters: the two Hall viscosities $\tilde{\eta}_\perp, \tilde{\eta}_{\parallel}$, and one Hall conductivity $\tilde{\s}$, and seven dissipative parameters: the two electrical conductivities $\s_\perp, \s_\parallel$, the two shear viscosities $\eta_\perp, \eta_\parallel$, and four bulk viscosities $\z_1, \z_2, \eta_1, \eta_2$, of which only three are independent due to the Onsager relation $3\z_2 - 6\eta_1 - 2\eta_2 = 0$. We review the construction of the equilibrium generating functional up to first order in derivatives, and vary it with respect to the sources to obtain the associated energy-momentum tensor and the conserved current. In section \ref{beyondeq}, we go beyond the assumption of thermal equilibrium, and include fluctuations about an equilibrium configuration. We write down the constitutive relations including non-equilibrium corrections, highlighting the role of various transport parameters that appear at this order. Subsequently, in section \ref{linear}, we construct the linearized hydrodynamic equations, and compute the retarded two-point correlation functions of the hydrodynamic degrees of freedom about a state of equilibrium, making use of linear response theory. Finally, in section \ref{nonlinear}, we include nonlinear terms in the hydrodynamic constitutive relations, and compute the long-time tails in the correlation functions of the energy-momentum tensor. This requires us to make use of the symmetric correlation functions that follow from linear response theory. These can be obtained from the retarded correlation functions using the fluctuation-dissipation relation. We illustrate the dependence of the long-time tails on two of the many transport parameters that occur for the fluid on the anisotropic background: the shear viscosity $\eta_\parallel$ and the bulk viscosity $\z_1$. The paper ends with a discussion in section \ref{discussion}.\\

\textit{Notation and conventions.}---The metric has signature $(-+++)$. Spacetime indices are denoted by Greek letters $\m, \n, \ldots = (0,1,2,3)$. Spatial indices are denoted by Latin letters $i, j, \ldots = (1,2,3)$. Spatial vectors are denoted in boldface, e.g.\ ${\bf k, x}$ etc. Scalar product of spatial vectors is denoted in boldface, e.g.\ ${\bf k \cdot x}$, whereas the Lorentz invariant inner product of four-vectors is denoted without boldface, e.g.\ $A \cdot B \equiv \eta_{\m\n} A^\m B^\n$. The Levi-Civita tensor density is given by
\bes
\e^{\m\n\r\s} = \f{\ve^{\m\n\r\s}}{\sqrt{-g}}, \, \, \e_{\m\n\r\s} = \sqrt{-g} \, \ve_{\m\n\r\s}, 
\ees
 with $\ve^{\m\n\r\s}$ being the ordinary Levi-Civita symbol, with $\ve^{0123} = 1$. An important identity we make use of is the contraction identity,
\bes
\e^{\m_1\ldots\m_p \n_1\ldots\n_m} \e_{\m_1\ldots\m_p \l_1\ldots\l_m} = (- 1)^n \, p! \, \d^{\n_1\ldots\n_m}_{\l_1\ldots\l_m},
\ees
where $n$ is the number of minus signs in the metric signature ($n = 1$ for us), and $p$ is the number of contracted indices. $\d^{\n_1\ldots\n_m}_{\l_1\ldots\l_m}$ is the generalized Kronecker delta function, which can be expressed as a determinant via
\begin{equation*}
\begin{split}
\d^{\n_1\ldots\n_m}_{\l_1\ldots\l_m} = 
\begin{vmatrix}
\d^{\n_1}_{\l_1} & \d^{\n_1}_{\l_2} & \cdots & \d^{\n_1}_{\l_m}  \\
\d^{\n_2}_{\l_1} & \d^{\n_2}_{\l_2} & \cdots & \d^{\n_2}_{\l_m} \\
\vdots & \vdots & \ddots & \vdots\\
\d^{\n_m}_{\l_1} & \d^{\n_m}_{\l_2} & \cdots & \d^{\n_m}_{\l_m} 
\end{vmatrix}.
\end{split}
\end{equation*}

\section{Thermal equilibrium in a strong magnetic field}
\label{thermodynamics}
When a system in thermal equilibrium is perturbed, it relaxes back to equilibrium via dynamical processes that are generically governed by two types of variables: slow and fast. The slow variables are typically associated with conserved quantities of the system, that can relax back only by physical transport, and thus govern the dynamics of the perturbed system over long time scales. The fast variables, on the other hand, are associated with non-conserved quantities, and can relax back to the new equilibrium state very quickly. 
Hydrodynamics captures the long-time low-energy behaviour of conserved quantities of the system, as the system relaxes back to thermal equilibrium after being perturbed. 

Before jumping into hydrodynamics, we would like to better understand the state of thermal equilibrium itself. Mathematically, thermal equilibrium is quantified via the presence of a timelike vector $V^\m$, which in the fluid rest frame takes the form $V^\m_{\rm rest} = (1, {\bf 0})$. The intuitive fact that things in equilibrium do not change with time is imposed by the condition that the Lie derivative of the hydrodynamic variables as well as the sources with respect to $V^\m$ vanishes. In fact, in thermal equilibrium, the hydrodynamic variables, which are the fluid four-velocity $u^\m$, temperature $T$, and chemical potential $\m$, can be defined in terms of the vector $V^\m$ and the sources as \cite{Banerjee:2012iz, Jensen:2012jh}
\be
\label{thermoframe}
T = \f{T_0}{\sqrt{-V^2}}\, , \quad u^\m = \f{V^\m}{\sqrt{-V^2}}\, , \quad \m = \f{V^\m A_\m + \L_V}{\sqrt{-V^2}}\, ,
\ee
where $V^2 \equiv g_{\m\n} V^\m V^\n$ is the squared norm of $V^\m$. The metric $g_{\m\n}$ and the vector field $A_\m$ are external sources, which respectively source the energy-momentum tensor and the conserved $U(1)$ current for the system. The normalization of the temperature is set by $T_0$, and $\L_V$ ensures that the chemical potential is gauge invariant \cite{Jensen:2013kka}. The fluid velocity is normalized such that $u_\m u^\m = -1$. The choice eq.\ \eqref{thermoframe} for defining the hydrodynamic variables is called the \textit{thermodynamic frame}.

We will be interested in a charged fluid with a global $U(1)$ symmetry kept in the presence of a strong background magnetic field. By the background magnetic field being strong, we imply that various susceptibilities and transport parameters of the fluid will depend not just upon the temperature and the chemical potential, but also on the strength of the magnetic field, as discussed further below.  Given the fluid four-velocity $u^\m$, one can express the antisymmetric field strength tensor $F_{\m\n}$ as
\be
\label{fieldstrength}
F_{\m\n} = u_\m E_\n - u_\n E_\m - \e_{\m\n\rho\s} u^\rho B^\s,
\ee
where the electric and magnetic fields are defined via $E_\m \equiv F_{\m\n} u^\n$, $B^\m = \half \e^{\m\n\rho\s} u_\n F_{\rho\s}$. Note that both the electric and magnetic fields are orthogonal to the fluid velocity, $u\cdot E = u \cdot B = 0$. Electric and magnetic fields are not independent, but linked to each other via the Bianchi identity $\e^{\m\n\r\s} \N_\n F_{\r\s} = 0$, which in thermal equilibrium becomes \cite{Hernandez:2017mch}
\begin{subequations}
\label{bianchi}
\begin{align}
&\N \cdot B = B \cdot a - E \cdot \O,\\
&u_\m \e^{\m\n\r\s}  \N_\r E_\s = u_\m \e^{\m\n\r\s} E_\r a_\s,
\end{align}
\end{subequations}
where $a^\m = u^\n \N_\n u^\m$ is the acceleration and $\O^\m = \e^{\m\n\r\s} u_\n \N_\r u_\s$ is the vorticity of the fluid. Note that eqs.\ \eqref{bianchi} are the curved space analogues of the flat space equilibrium identities ${{\bf\N \cdot B} = 0, {\bf \N \times E}= 0}$.

In thermal equilibrium, the correlation functions of conserved currents can be obtained from an equilibrium generating functional $\mc{W}[g_{\m\n}, A_{\m}]$, which is a functional of the external sources \cite{Banerjee:2012iz, Jensen:2012jh}.\footnote{As an aside, the equilibrium generating functional can be promoted to be the equilibrium effective action, and can be used to derive effective Einstein's equations incorporating the effects of derivative corrections beyond the perfect fluid approximation \cite{Kovtun:2019wjz}. See \cite{Kovtun:2016lfw} for a similar analysis for electromagnetism.} The variation of $\mc{W}$ with respect to the sources gives
\be
\d \mc{W} = \half \int d^4x \sqrt{-g} \, T^{\m\n} \d g_{\m\n} + \int d^4x \sqrt{-g}\, J^\m \d A_\m,
\ee
which implies that the one-point functions of the conserved currents are
\be
\label{onepoint}
T^{\m\n} = \f{2}{\sqrt{-g}} \f{\d \mc{W}}{\d g_{\m\n}}\, , \quad J^\m = \f{1}{\sqrt{-g}} \f{\d \mc{W}}{\d A_{\m}}.
\ee
Higher point correlation functions in equilibrium can be obtained by taking further derivatives with respect to the sources. The conservation equations 
\begin{subequations}
\label{hydroeqs}
\begin{align}
&\N_\m T^{\m\n} = F^{\n\r} J_\r, \\
&\N_\m J^\m = 0,
\end{align}
\end{subequations}
follow from the diffeomorphism and gauge invariance of the generating functional. The generating functional can be expressed as an integral over a local free energy density $\mc{F}$ via
\be
\label{defgenfunc}
\mc{W}[g_{\m\n}, A_\m] = \int d^4x \sqrt{-g} \, \mc{F}(g_{\m\n}, A_\m).
\ee
When the sources vary on length scales much longer than the  inherent microscopic scales in the system, such as the correlation length, the density $\mc{F}$ admits an expansion in terms of the derivatives of the hydrodynamic variables and the sources \cite{Banerjee:2012iz, Jensen:2012jh}. At zeroth order in the derivative expansion, for instance, we have
\be
\label{p1}
\mc{F}^{(0)} = p(T, \m, B^2),
\ee
where $p$ is the equilibrium pressure.

An important point here is the choice of the derivative counting scheme. The metric is naturally counted as $\mc{O}(1)$. For conducting fluids in the presence of strong magnetic fields, the appropriate derivative counting scheme is $B^\m \sim \mc{O}(1)$. Electric field in equilibrium satisfies the condition $\del_\l \m = E_\l - \m a_\l$, and hence it is appropriate to consider $E^\m \sim \mc{O}(\del)$, which is another way to say that the electric field is screened in charged fluids. Since the magnetic field is $\mc{O}(1)$, the thermodynamic susceptibilities and transport parameters of the fluid will depend upon its strength as well. For instance, the equation of state in eq.\ \eqref{p1} has an explicit dependence on $B^2$.

\subsection{The equilibrium constitutive relations}
The energy-momentum tensor for the fluid can be decomposed into components along and orthogonal to the fluid four-velocity $u^\m$ via
\be
\label{constT}
T^{\m\n} = \mc{E} u^\m u^\n + \mc{P} \D^{\m\n} + \mc{Q}^\m u^\n +  \mc{Q}^\n u^\m + \mc{T}^{\m\n},
\ee
where $\D_{\m\n} = g_{\m\n} + u_\m u_\n$ is the projector orthogonal to the fluid velocity, the heat current $\mc{Q}^\m$ is orthogonal to the fluid velocity, and $\mc{T}^{\m\n}$ is symmetric, traceless and transverse to the fluid velocity. Given $T^{\m\n}$ one can extract the coefficients via
\be
\begin{split}
&\mc{E} = u_\m u_\n T^{\m\n}, \,\, \mc{P} = \f{1}{3} \D_{\m\n}T^{\m\n}, \,\, \mc{Q}_\m = - \D_{\m\n}u_\r T^{\n\r}\\
&\mc{T}_{\m\n} = \f{1}{2} \left(\D_{\m\a} \D_{\n\b} + \D_{\m\b} \D_{\n\a}  - \f{2}{3} \D_{\m\n} \D_{\a\b} \right) T^{\a\b}.
\end{split}
\ee
In the hydrodynamic regime, the quantities $\mc{E}, \mc{P}, \mc{Q}^\m, \mc{T}^{\m\n}$ can be expanded in a series in the number of derivatives acting on the hydrodynamic variables and the sources, which are called the \textit{constitutive relations} for the fluid. Just like the energy-momentum tensor, the conserved $U(1)$ current can also be decomposed as
\be
\label{constJ}
J^\m = \mc{N} u^\m + \mc{J}^\m,
\ee
where $\mc{J}^\m$ is transverse to the fluid velocity. Given the current $J^\m$, one can extract the coefficients in eq. \eqref{constJ} via
\be
\label{constJ2}
\mc{N} = - u_\m J^\m, \, \, \mc{J}_\m = \D_{\m\n} J^\n.
\ee

\subsubsection{The zeroth order}
As an illustration, let us compute the constitutive relations for the fluid at the zeroth order in the derivative expansion \cite{Hernandez:2017mch}. The equilibrium generating functional at the zeroth order in derivatives is given by
\be
\mc{W}[g_{\m\n}, A_\m] = \int d^4x \sqrt{-g}\, p(T,\m,B^2). 
\ee
Varying it with respect to the sources $g_{\m\n}$ and $A_{\m}$ gives the zeroth order constitutive relations,
\begin{subequations}
\label{constrel1}
\begin{align}
&\mc{E} = - p + T \f{\del p}{\del T} + \m \f{\del p}{\del \m} \equiv \e(T, \m, B^2), \label{defeps}\\
&\mc{P} = p - \f{2}{3} \a_{\rm BB} B^2, \label{defPi}\\
&\mc{N} = \f{\del p}{\del \m} \equiv n(T, \m, B^2), \\
&\mc{T}^{\m\n} = \a_{BB} \left(B^\m B^\n - \f{1}{3} \D^{\m\n} B^2 \right),
\end{align}
\end{subequations}
where $\e(T,\m,B^2)$ is the equilibrium energy density and $n(T,\m,B^2)$ is the equilibrium charge density. 
Also, $\a_{\rm BB}(T, \m, B^2) = 2 \del p/\del B^2$ denotes the magnetic susceptibility. At the zeroth order, one has $\mc{Q}^\m = \mc{J}^\m = 0$.

\subsubsection{The first order}
For a parity violating fluid, at the first order in derivatives in thermal equilibrium, there are five independent scalars that can contribute to the generating functional $\mc{W}[g_{\m\n}, A_\m]$, which consequently takes the form \cite{Hernandez:2017mch}\footnote{Note that when the magnetic field is weak, $B^\m \sim \mc{O}(\del)$, derivative contributions to the equilibrium generating functional begin at the second order \cite{Kovtun:2018dvd}. See also \cite{Shukla:2019shf} for the behaviour of  second order susceptibilities for free quantum field theories as a function of $\m, T$.}
\be
\mc{W} = \int d^4x \sqrt{-g} \left(p(T,\m,B^2) + \sum_{i=1}^5 M_i(T, \m, B^2) \, S_i \right),
\ee
where the one-derivative scalars $S_i$ are
\be
\begin{split}
&S_1 = B^\m \del_\m \left( \f{B^2}{T^4} \right), \,\,  S_2 = \e^{\m\n\r\s} u_\m B_\n \N_\r B_\s, \\
&S_3 = B\cdot a,\,\, S_4 = B\cdot \O, \,\, S_5 = B\cdot E.
\end{split}
\ee
Out of these, only $S_4$ is parity preserving, and the rest are parity violating. Also, $M_i(T,\m,B^2)$ are thermodynamic susceptibilities and need to be determined from the microscopic theory, just like the pressure $p$.

Our focus in the present work is on parity preserving fluids in 3+1D.\footnote{See \cite{Abbasi:2015saa, Abbasi:2016rds, Ammon:2020rvg} for computations in 3+1D where parity violation is considered. Additionally, see \cite{Hartnoll:2007ih, Baggioli:2020edn, Amoretti:2020mkp, Amoretti:2021fch} for the effects of strong background magnetic fields in 2+1D.} Thus, for our purposes, the equilibrium generating functional up to first order in derivatives is given by (with the susceptibility $M_4$ denoted more suggestively by $M_\O$)
\be
\mc{W} = \int d^4x \sqrt{-g}\, \left [p(T,\m,B^2) + M_\O \, B\cdot \O\right].
\label{genfuncfir}
\ee
Varying this with respect to the metric $g_{\m\n}$ gives us the following first order constitutive relations for the energy-momentum tensor eq.\ \eqref{constT},\footnote{The terms in the last line of eq.\ \eqref{constTfir4} have extra factors of 2 when compared with the results in \cite{Hernandez:2017mch}. This is so because the authors of \cite{Hernandez:2017mch} use two different definitions for the projection $A^{\langle\m} B^{\n\rangle}$, whereas we uniformly use the definition given in eq.\ \eqref{defang}.}
\begin{subequations}
\label{constTfir}
\begin{align}
&\mc{E} = \e + \left( T M_{\O,T} + \m M_{\O,\m} - 2M_\O\right) B\cdot \O\, ,\label{constTfir1}\\
&\mc{P} = p-\f{2}{3} \a_{\rm BB} B^2 - \f{1}{3} \left(M_\O + 4 B^2 M_{\O,B^2}\right) B\cdot \O\, , \label{constTfir2}\\
&\mc{Q}^\m = M_\O \e^{\m\n\r\s} u_\n \del_\r B_\s - M_{\O, B^2} \e^{\m\n\r\s} u_\n B_\r \del_\s B^2 + \left(2M_\O -T M_{\O,T} - \m M_{\O,\m}\right) \e^{\m\n\r\s} B_\n u_\r a_\s \nonumber\\
&\quad\quad  + \left(M_{\O,\m} - 2M_{\O,B^2} B\cdot \O - \a_{BB}\right)  \e^{\m\n\r\s} u_\n E_\r B_\s +  M_\O \e^{\m\n\r\s} \O_\n E_\r u_\s\, ,\label{constTfir3}\\
&\mc{T}^{\m\n} = \a_{\rm BB}\left(B^\m B^\n - \f{1}{3} \D^{\m\n} B^2\right) + 2 M_{\O,B^2} B\cdot \O \, B^{\langle \m} B^{\n\rangle} + 2 M_\O B^{\langle \m} \O^{\n\rangle}\, \label{constTfir4},
\end{align}
\end{subequations}
with $\a_{\rm BB} = 2 \del p/\del B^2$ as defined earlier. A comma in the subscript denotes a partial derivative with respect to the argument that follows. In deriving the  constitutive relations in eqs.\ \eqref{constrel1} and \eqref{constTfir}, we made use of the following intermediate results for the variation of various quantities that enter the generating functional under a variation of the metric. These are
\begin{subequations}
\begin{align}
&\d T = \f{T}{2}\, u^\m u^\n \d g_{\m\n}\, , \\
&\d \m = \f{\m}{2}\, u^\m u^\n \d g_{\m\n}\, , \\
&\d B^\m = - \half B^\m \D^{\a\b} \d g_{\a\b} + u^\m u^\a B^\b \d g_{\a\b} + \e^{\m\a\r\s} u^\b u_\r E_\s \d g_{\a\b}\, ,\\
&\d B^2 = B^\m B^\n \d g_{\m\n} - B^2 \D^{\m\n} \d g_{\m\n} - 2 \e^{\m\n\r\s} u_\n E_\r B_\s u^\a \d g_{\m\a}\, ,\\
&\d\O^\m = \O^\m \left(u^\a u^\b - \half g^{\a\b}\right) \d g_{\a\b} + \e^{\m\n\r\s} u^\a \N_\r u_\s \d g_{\n\a} + \e^{\m\n\r\s} u_\n \del_\r u^\a \d g_{\a\s} + \e^{\m\n\r\s} u_\n u^\a \del_\r\d g_{\a\s}\\
&\delta(B\cdot \O) = B^\m \O^\n \d g_{\m\n} + \left(\half u^\m u^\n - g^{\m\n}\right) B\cdot\O \, \d g_{\m\n} \nonumber\\
&\qquad\qquad- \e^{\m\a\b\r} B_\a \N_\b u_\r\, u^\n  \d g_{\m\n} - \e^{\m\a\b\r} B_\a u_\b \del_\r u^\n  \d g_{\m\n} - \e^{\m\a\b\r} B_\a u_\b u^\n \del_\r \d g_{\m\n}.
\end{align}
\end{subequations}
Also, the angular bracket notation used above helps in defining the symmetric transverse traceless part of a tensor, given by
\be
A^{\langle\m} B^{\n\rangle} \equiv \half \left(\D^{\m\a} \D^{\n\b} + \D^{\m\b} \D^{\n\a} - \frac{2}{3} \D^{\m\n} \D^{\a\b}   \right) A_\a B_\b.
\label{defang}
\ee

For computing the constitutive relations for the $U(1)$ current, it is convenient to regard the generating functional as a functional of $A_\m$ and its derivative $F_{\m\n}$. The variation of the generating functional is then defined as \cite{Kovtun:2016lfw, Kovtun:2018dvd}
\be
\d_{A,F} \mc{W} = \int d^4x \sqrt{-g} \left( J^\m_f \, \d A_{\m} + \half M^{\m\n} \d F_{\m\n}\right)\, ,
\ee
where $J^\m_f$ is the current of free charges and $M^{\m\n}$ is the antisymmetric polarization tensor. In terms of these the total current can be written as
\be
\label{currdec}
J^\m = J^\m_f - \N_\l M^{\l\m}. 
\ee
The ambiguity in defining the free charge current is fixed by choosing $J^\m_f = \rho u^\m$, with $\rho = \del \mc{F}/\del \m$ being the free charge density. Comparing eqs.\ \eqref{constJ}, \eqref{constJ2} with eq.\ \eqref{currdec} above, we find the constitutive relations for the $U(1)$ current to be
\begin{subequations}
\label{constJpol}
\begin{align}
&\mc{N} = \r - \N \cdot p + p\cdot a - m\cdot \O \, ,\\
&\mc{J}^\m = \e^{\m\n\r\s} u_\n \left( \N_\r + a_\r \right) m_\s\, ,
\end{align}
\end{subequations}
where we have made use of the electric and magnetic polarization vectors, denoted respectively by $p^\m$ and $m^\m$, in writing the constitutive relations eqs.\ \eqref{constJpol}, with
\be
p^\m = u_\n M^{\n\m}\, , \quad  m^\m = \half \e^{\m\n\r\s} u_\n M_{\r\s}.
\ee

For the first order generating functional eq.\ \eqref{genfuncfir} we find the polarization vectors
\begin{subequations}
\begin{align}
&p^\m = 0\, ,\\
&m^\m = \left( \a_{\rm BB} + 2 M_{\O,B^2} B\cdot \O\right) B^\m + M_\O \O^\m\, ,
\end{align}
\end{subequations}
which lead to the constitutive relations
\begin{subequations}
\label{constJfir}
\begin{align}
&\mc{N} = n + \f{\del M_\O}{\del \m} B\cdot \O - m\cdot \O\, , \label{eqN}\\
&\mc{J}^\m = \e^{\m\n\r\s} u_\n \left( \N_\r + a_\r \right) m_\s. \label{eqJ}
\end{align}
\end{subequations}

\section{Beyond thermal equilibrium}
\label{beyondeq}
Let us now move on and introduce deviations from the state of thermal equilibrium introduced in section \ref{thermodynamics}. A comprehensive analysis of the out-of-equilibrium constitutive relations in the presence of a strong magnetic field appears in \cite{Hernandez:2017mch}, which we shall make use of. It is important to notice that hydrodynamic variables such as the temperature, chemical potential or the fluid velocity have no a priori definition once one considers a state out of equilibrium. In other words, hydrodynamic variables are well defined only for a system in equilibrium. The freedom to choose a definition for hydrodynamic variables out of equilibrium, called a choice of hydrodynamic frame, can be used to fix some of the constitutive relations to a simple form. Following \cite{Hernandez:2017mch}, we choose to work in the thermodynamic frame. In this frame, the freedom to choose the temperature $T$ and the chemical potential $\m$ out of equilibrium can be used to fix $\mc{E}$ and $\mc{N}$ to their equilibrium values, given by eqs.\ \eqref{constTfir1} and \eqref{eqN}. Also, the freedom to redefine the fluid velocity can be used to fix $\mc{Q}^\m$ to its equilibrium value \eqref{constTfir3}. The other three quantities $\mc{P}, \mc{J}^\m$ and $\mc{T}^{\m\n}$ in the constitutive relations receive derivative corrections beyond their equilibrium values. The thermodynamic frame constitutive relations for $\mc{P}, \mc{J}^\m$ and $\mc{T}^{\m\n}$ with first-order derivative corrections out of equilibrium are given by
\begin{subequations}
\label{constfirst}
\begin{align}
&\mc{P} = \bar{\mc{P}} - \z_1 \N\cdot u - \z_2 \, b^\m b^\n \N_\m u_\n\, ,\\
&\mc{T}^{\m\n} = \bar{\mc{T}}^{\m\n} - \eta_\perp \s^{\m\n}_{\perp} - \eta_\parallel (b^\m \S^\n + b^\n \S^\m) - b^{\langle \m} b^{\n\rangle} (\eta_1 \N\cdot u + \eta_2 b^\a b^\b \N_\a u_\b) \nonumber\\
&\quad\quad- \tilde{\eta}_\perp \tilde{\s}^{\m\n}_\perp - \tilde{\eta}_\parallel (b^\m \tilde{\S}^\n + b^\n \tilde{\S}^\m)\, , \\
&\mc{J}^\m = \bar{\mc{J}}^\m + \left(\s_\perp \mathbb{B}^{\m\n} + \s_\parallel \f{B^\m B^\n}{B^2}\right) \mathbb{V}_\n + \tilde{\s} \tilde{\mathbb{V}}^\m.
\end{align}
\end{subequations}
Here $\bar{\mc{P}}, \bar{\mc{J}}^\m, \bar{\mc{T}}^{\m\n}$ are the equilibrium values, given in eqs.\ \eqref{constTfir2}, \eqref{eqJ} and \eqref{constTfir4}, respectively. Also, we have used the notation $b^\m \equiv B^\m/\sqrt{B^2}$, and $\mathbb{B}^{\m\n} \equiv \D^{\m\n} - \f{B^\m B^\n}{B^2}$ is the projector orthogonal to both $u^\m$ and $B^\m$. Also, we use the shorthand notation $\mathbb{V}^\m \equiv E^\m - T \D^{\m\n} \del_\n\left(\frac{\m}{T}\right)$, and $\tilde{\mathbb{V}}^\m = \e^{\m\n\a\b} u_\n b_\a \mathbb{V}_\b$. A decomposition similar to eq.\ \eqref{constfirst} for the first-order out-of-equilibrium constitutive relations in the Landau frame appears in \cite{Huang:2011dc}. 

The shear tensor $\s^{\m\n}$, given by $$\s^{\m\n} = \D^{\m\a}  \D^{\n\b} \left( \N_\a u_\b + \N_\b u_\a - \f{2}{3}\, g_{\a\b} \N\cdot u\right),$$ has been decomposed as
\be
\s^{\m\n} = \s^{\m\n}_\perp + \left(b^\m \S^\n + b^\n \S^\m \right) + \half b^{\langle \m} b^{\n \rangle} (3 b^\a b^\b \N_\a u_\b - \N\cdot u).
\label{sigdec1}
\ee
Here $\s^{\m\n}_\perp$ is transverse to both $u^\m$ and $B^\m$, $$\s^{\m\n}_\perp = \half \left( \mathbb{B}^{\m\a} \mathbb{B}^{\n\b} + \mathbb{B}^{\m\b} \mathbb{B}^{\n\a} - \mathbb{B}^{\m\n} \mathbb{B}^{\a\b}\right)\s_{\a\b}\, .$$ Similarly, $\S^\m \equiv \mathbb{B}^{\m\a} \s_{\a\b} b^\b$ is also transverse to both $u^\m$ and $B^\m$.

The tensor $\tilde{\s}^{\m\n}$ is another rank-two one-derivative tensor that enters the constitutive relations above. It is given by
\be
\tilde{\s}^{\m\n} = \f{1}{2\sqrt{B^2}} \left( \e^{\m\r\a\b} u_\r B_\a \s_\b^{\,\,\n} + \e^{\n\r\a\b} u_\r B_\a \s_\b^{\,\,\m}\right).
\label{deftils}
\ee
Inserting the decomposition eq.\ \eqref{sigdec1} for $\s^{\m\n}$ into eq.\ \eqref{deftils}, one gets
\be
\tilde{\s}^{\m\n} = \tilde{\s}^{\m\n}_\perp + \left( b^\m \tilde{\S}^\n + b^\n \tilde{\S}^\m\right),
\ee
where $$\tilde{\s}^{\m\n}_\perp = \half \left( \mathbb{B}^{\m\a} \mathbb{B}^{\n\b} + \mathbb{B}^{\m\b} \mathbb{B}^{\n\a} - \mathbb{B}^{\m\n} \mathbb{B}^{\a\b}\right) \tilde{\s}_{\a\b}\, ,$$ and $\tilde{\S}^\m = \mathbb{B}^{\m\a} \tilde{\s}_{\a\b} b^\b$.

The first-order out-of-equilibrium transport parameters appearing in the constitutive relations eq.\ \eqref{constfirst} are not all dissipative. In general, every dissipative transport parameter is necessarily an out-of-equilibrium contribution, but every out-of-equilibrium parameter is not necessarily dissipative. The eleven transport parameters appearing in eq.\ \eqref{constfirst} can be classified as follows.

\begin{center}
\begin{tabular}{|c||c||c|} 
 \hline
\textit{Parameter} & \textit{Name} & \textit{Nature} \\ 
\hline
\hline
$\eta_\perp$  & Transverse shear viscosity & Dissipative \\ 
\hline
$\eta_\parallel$ & Longitudinal shear viscosity & Dissipative \\ 
\hline
$\s_\perp$ & Transverse conductivity& Dissipative\\
\hline
$\s_\parallel$ & Longitudinal conductivity & Dissipative\\
\hline
$\z_1, \z_2, \eta_1, \eta_2$ & Bulk viscosities & Dissipative\\
\hline
$\tilde{\eta}_\perp$  & Transverse Hall viscosity & Non-diss. \\ 
\hline
$\tilde{\eta}_\parallel$ & Longitudinal Hall viscosity & Non-diss. \\ 
\hline
$\tilde{\s}$ & Hall conductivity & Non-diss.\\
\hline
\end{tabular}
\end{center}

An important point to note is that not all of the four bulk viscosities are independent. The time reversal invariance of the underlying microscopic theory imposes an Onsager relation on the bulk viscosities above, and gives the constraint 
\be
3\z_2 - 6\eta_1 - 2\eta_2 = 0.
\ee
Thus, the number of independent out-of-equilibrium transport parameters at one-derivative order is ten. Also, if one assumes the fluid to be conformal, then the bulk viscosities $\z_1 = \z_2 =0$.

\section{Linearized hydrodynamics}
\label{linear}
We now move on to study the behaviour of small fluctuations and their correlation functions about a state of thermal equilibrium, which is the main thrust of this paper. The equilibrium state we consider coincides with the fluid at rest $u^\m_0 = (1,\v[0])$, at a constant global temperature $T_0$ and chemical potential $\mu_0$. The equilibrium energy density is $\e_0$, pressure $p_0$ and charge density $n_0$. The background magnetic field can be taken to point along the $+Z$ direction without any loss of generality, so $B^\m = (0,0,0, B_0)$. All other background sources are turned off. We work with fluctuations in the energy density $\d\e$, charge density $\d n$, and momentum density $\pi^i$. These are defined via
\be
\begin{split}
&T^{00} = \e_0 + \d\e + \ldots\, ,\\
&T^{0i} = \pi^i + \ldots\, ,\\
&J^0 = n_0 + \d n + \ldots\, ,
\end{split}
\ee
where the $\ldots$ in the expressions above denote terms which depend upon the background magnetic field $B_0$. Note that the momentum density fluctuations are related to the velocity fluctuations via $\pi^i = (\e_0 + p_0) \d u^i = w_0 \d u^i$, where $w_0 = \e_0 + p_0$ is the equilibrium enthalpy density.

Using the thermodynamic frame constitutive relations for the energy-momentum tensor and the conserved current discussed in sections \ref{thermodynamics} and \ref{beyondeq}, one can write down the hydrodynamic equations \eqref{hydroeqs} to linear order in the fluctuations $\d n, \d\e, \pi^i$ about the chosen state of equilibrium. The full set of linearized equations about the equilibrium state of interest are presented in \ref{full_lin}.

The primary reason to work out the linearized fluctuation equations is to be able to make use of linear response theory, which would allow us to obtain retarded correlation functions of the fluctuations about the given state of equilibrium. As per the linear response approach, for a set of hydrodynamic variables $\Phi_{\rm A}(t,{\bf x})$ satisfying coupled linear equations in momentum space of the form
\be
\del_t \Phi_{\rm A} (t,{\bf{k}}) + \mb{M}_{\rm AB}({\bf k}) \, \Phi_{\rm B}(t, {\bf k}) = 0,
\ee
the retarded two-point correlation functions are given by\footnote{The retarded and symmetric correlation functions are defined via
\begin{align*}
&\mb{G}^{\rm Ret}_{\mc{O}_1 \mc{O}_2} (t_1-t_2, {\bf x}_1 -{\bf x}_2) \equiv -i \Theta(t_1-t_2) \langle[\mc{O}_1(t_1, {\bf x}_1), \mc{O}_2(t_2, {\bf x}_2)] \rangle ,\\
&\mb{G}^{\rm Sym}_{\mc{O}_1 \mc{O}_2} (t_1-t_2, {\bf x}_1 -{\bf x}_2) \equiv \half  \langle\{\mc{O}_1(t_1, {\bf x}_1), \mc{O}_2(t_2, {\bf x}_2)\} \rangle ,
\end{align*}
where $\Theta(t)$ is the step function, and square and curly braces respectively denote a commutator and an anticommutator.}
\be
\label{techno}
\mb{G}^{\rm Ret}(\o, {\bf k}) =   - \left(\boldsymbol{1} + i \o \mb{K}^{-1}\right) \boldsymbol{\chi},
\ee
where the matrix $\mb{K} = - i \o \boldsymbol{1} + \mb{M}({\bf k})$. Also, if the sources for the hydrodynamic fluctuations $\Phi_A$ are $\phi_A$, then the susceptibility matrix $\boldsymbol{\chi}$ is given by 
\be
\boldsymbol{\chi}_{\rm AB} = \f{\del \Phi_A}{\del \phi_B}.
\label{defsusc}
\ee
The matrix $\boldsymbol{\chi}$ has been computed in \ref{suscmat}. Linear response theory is thus good enough if one is interested only in the causal linear response of the system to external sources. However, the richness of hydrodynamics is contained in the nonlinear nature of its equations, and we explore some of their consequences in section \ref{nonlinear} below.

Making use of the linear response approach outlined above, in principle one can work with the entire set of transport parameters that appear at the first-order and compute the linear response of the system following from the hydrodynamic equations in \ref{full_lin}. However, this is quite cumbersome and not analytically tractable. A simpler strategy is to compute the linear response when one of the first-order transport parameters dominates over the rest. In such a scenario, one can assume that it is only this transport parameter that is non-zero, while the rest are set to vanish. This approximation of dealing with a single transport parameter and its effects on the system at a given time renders the setup analytically tractable. Needless to say, one can work with all or any other suitable subset of non-zero transport parameters depending upon the situation of interest. However, this will inevitably require the use of numerical approaches to the problem, which we leave out for a future study.

In section \ref{nonlinear} below, we will illustrate the effect of non-linearities and the emergence of long-time tails for the two cases when the dissipative viscosities $\eta_\parallel$ and  $\zeta_1$ are the ones that dominate the transport and relaxation process. We make this choice for the relative simplicity of the equations and loop-computations that follow. Further, it is useful to assume that we are working with a  fluid which is perturbed around a neutral state i.e.\ $n_0 = 0$. With this in mind, we now move on to compute the linear response and the associated retarded correlation functions for the two cases of interest.

\subsection{Linear response when only $\eta_\parallel \neq 0$}
\label{linrep1}
We first consider the case when the dominant transport parameter is the shear viscosity $\eta_\parallel$, and the other transport parameters are negligibly small. Since we are interested in the shear modes, an additional simplification one can impose is the assumption of incompressibility, $\del_i \pi^i \approx 0$, which decouples the sound modes of the system and further simplifies the coupled equations in \ref{full_lin}. In fact, with these simplifications put in, the energy density and charge density fluctuations decouple from the velocity fluctuations, leaving behind a much simpler system of equations to be analyzed for the shear modes.
Linear response theory then leads to the following set of retarded two-point correlators. 
\begin{subequations}
\begin{align}
&\mb{G}^{\rm Ret}_{\pi_x \pi_x}(\o, {\bf k}) = \f{\eta_\parallel k_\parallel^2 \left[i\o- \f{\eta_\parallel}{w_0} (k^2 -k_x^2)\right]}{{F}_1(\o, k_\parallel)\, F_2(\o, k)} , \label{etaxx}\\
&\mb{G}^{\rm Ret}_{\pi_y \pi_y}(\o, {\bf k}) = \f{\eta_\parallel k_\parallel^2 \left[i\o- \f{\eta_\parallel}{w_0} (k^2 -k_y^2)\right]}{F_1(\o, k_\parallel) \, F_2(\o, k)} ,\\
&\mb{G}^{\rm Ret}_{\pi_z \pi_z}(\o, {\bf k}) = \f{\eta_\parallel k_\perp^2}{F_2(\o, k)} ,\label{etazz}\\
&\mb{G}^{\rm Ret}_{\pi_x \pi_y}(\o, {\bf k}) = \f{\eta_\parallel^2 k_\parallel^2 k_x k_y}{w_0 F_1(\o, k_\parallel) \, F_2(\o, k)} ,\\
&\mb{G}^{\rm Ret}_{\pi_x \pi_z}(\o, {\bf k}) = \f{\eta_\parallel k_x k_z}{F_2(\o, k)} ,\\
&\mb{G}^{\rm Ret}_{\pi_y \pi_z}(\o, {\bf k}) = \f{\eta_\parallel k_y k_z}{F_2(\o, k)} ,
\end{align}
\end{subequations}
where 
\begin{subequations}
\begin{align*}
&{F}_1(\o, k_\parallel) = i \o - \frac{\eta_\parallel}{w_0} k_\parallel^2 ,\\
&F_2(\o, k) =  i \o - \frac{\eta_\parallel}{w_0} k^2.
\end{align*}
\end{subequations}
Note that since the magnetic field points along the $+Z$ direction, in the above we have made use of the intuitive notation $k_\parallel \equiv k_z, k_\perp = \sqrt{k_x^2 + k_y^2},\, k \equiv |{\bf k}|$.

\subsection{Linear response when only $\z_1 \neq 0$}
\label{linrep2}
This is the second of the cases we will consider, when the bulk viscosity $\z_1$ is the dominant transport parameter, while the rest are negligibly smaller. The set of non-vanishing retarded two-point functions for this case that arise from linear response theory are as follows.
\begin{subequations}
\begin{align}
&\mb{G}^{\rm ret}_{\e\e}(\o, {\bf k}) =  \f{w_0 k^2}{(\o-\o_+)(\o-\o_-)}\, ,\\ 
&\mb{G}^{\rm ret}_{\e\pi_i}(\o, {\bf k}) =  \f{w_0 \,\o\, k_i}{(\o-\o_+)(\o-\o_-)}\, ,\\ 
&\mb{G}^{\rm ret}_{\pi_x \pi_x}(\o, {\bf k}) = \f{k_x^2 \left(w_0 v_s^2 - i\o\z_1\right)}{(\o-\o_+)(\o-\o_-)}\, ,\\
&\mb{G}^{\rm ret}_{\pi_y \pi_y}(\o, {\bf k}) = \f{k_y^2 \left(w_0 v_s^2 - i\o\z_1\right)}{(\o-\o_+)(\o-\o_-)} \, ,\\
&\mb{G}^{\rm ret}_{\pi_z \pi_z}(\o, {\bf k}) = \f{k_z^2 \left(w_0 v_s^2 - i\o\z_1\right)}{(\o-\o_+)(\o-\o_-)}\, ,\\
&\mb{G}^{\rm ret}_{\pi_x \pi_y}(\o, {\bf k}) = \f{k_x k_y \left(w_0 v_s^2 - i\o\z_1\right)}{(\o-\o_+)(\o-\o_-)}\, ,\\
&\mb{G}^{\rm ret}_{\pi_x \pi_z}(\o, {\bf k}) = \f{k_x k_z \left(w_0 v_s^2 - i\o\z_1\right)}{(\o-\o_+)(\o-\o_-)}\, ,\\
&\mb{G}^{\rm ret}_{\pi_y \pi_z}(\o, {\bf k}) = \f{k_y k_z \left(w_0 v_s^2 - i\o\z_1\right)}{(\o-\o_+)(\o-\o_-)}\, ,
\end{align}
\end{subequations}
where $\o_\pm$ are the two sound modes, given by
$$\o_\pm = \pm v_s k - i \f{\z_1}{2w_0} k^2.$$
Note that $v_s$ is the speed of sound, which in the neutral equilibrium state of interest is given by $v_s^2 = (\del p/\del \e)$.

\section{Interactions and long-time tails}
\label{nonlinear}
With the retarded correlation functions in hand, we now move on to compute the long-time tails, and their dependence on the viscosities $\eta_\parallel, \z_1$. As we have mentioned earlier, the long-time tails are a consequence of the nonlinear terms in the hydrodynamic constitutive relations, which correspond to interactions between the fluctuating degrees of freedom. The leading long-time tails will come from the leading nonlinear terms in the energy-momentum tensor eq.\ \eqref{constTfir}, which have the form
\begin{align}
&T^{ij} = \d^{ij} \left(\Upsilon_1 \,\d\e^2 + \Upsilon_2\, \d n^2-\f{v_s^2}{w_0}{\boldsymbol \pi}^2 + \half \f{\a_{\rm BB} B_0^2}{w_0^2} \pi_z^2\right) + \left(w_0 - \a_{\rm BB} B_0^2\right) \f{\pi^i \pi^j}{w_0^2}  \nonumber\\
&\qquad+\left(\a_{BB} - M_{\O,\m} \right) \f{B_0^2}{w_0^2} 
\left[\pi_x(\d^{ix}\pi^j + \d^{jx} \pi^i) + \pi_y(\d^{iy}\pi^j + \d^{jy} \pi^i) \right]+  \cdots \label{stressnl}
\end{align}
where $\Upsilon_1 = \half \left(\frac{\del^2 p}{\del\e^2}\right)_n$, $\Upsilon_2 = \half \left(\frac{\del^2 p}{\del n^2}\right)_\e$. We have dropped nonlinear terms involving any derivatives, denoted by the $\cdots$, as their contribution in the hydrodynamic limit $\o \rightarrow 0, {\bf k} \rightarrow 0$ will be subleading as compared to the leading terms above. Terms higher than quadratic will not be relevant for the computation of the long-time tails, as we discuss below. 

Before we proceed to calculate the long-time tails for the cases of interest, let us intuitively explain their origin from the nonlinear interaction terms in hydrodynamic constitutive relations. Suppose we are interested in computing the effects that follow from quadratic fluctuation terms present in the nonlinear constitutive relation for a given conserved current $\mc{J}$. For concreteness, let us assume that the constitutive relation has the generic form $\mc{J}_{quad} \sim \Phi_1 \Phi_2$, where $\Phi_1$ and $\Phi_2$ are two fluctuating hydrodynamic degrees of freedom. Then the symmetric two-point correlation function of $\mc{J}$ receives a nontrivial contribution from the quadratic fluctuation terms, given by
\be
\label{fourpoint1}
{\rm G}^{\rm int}_{\mc{J} \mc{J}}(t, {\bf x}) \sim  \langle \Phi_1(t, {\bf x}) \Phi_2(t, {\bf x})  \Phi_1(0) \Phi_2(0)\rangle.
\ee
Now assuming that the small fluctuations about the state of equilibrium are Gaussian in nature, the four-point correlation function above factorizes into a product of two two-point correlation functions, and is given by
\be
\label{genform1}
{\rm G}^{\rm int}_{\mc{J} \mc{J}}(\o, {\bf k}) \sim \int \f{d\o'}{2\pi} \f{d^3k'}{(2\pi)^3} \,\mb{G}^{\rm sym}_{\Phi_1 \Phi_1}(\o', {\bf k}') \, \mb{G}^{\rm sym}_{\Phi_2 \Phi_2}(\o-\o', {\bf k} - {\bf k}')\, ,
\ee 
where $\mb{G}^{\rm sym}_{\mc{O}_1 \mc{O}_2}$ is the symmetric two-point correlation function that follows from linear response theory. These can be computed from the retarded two-point correlators of section \ref{linear} by using the fluctuation-dissipation relation,
\be
\label{fdt}
\mb{G}^{\rm Sym}_{\mc{O}_1 \mc{O}_2}(\o, {\bf k}) = - \f{2 T_0}{\o} \, {\rm Im} \, \mb{G}^{\rm Ret}_{\mc{O}_1 \mc{O}_2}(\o, {\bf k}).
\ee
Now the integral in eq.\ \eqref{genform1} is essentially a one-loop integral, as depicted in the diagram below. 
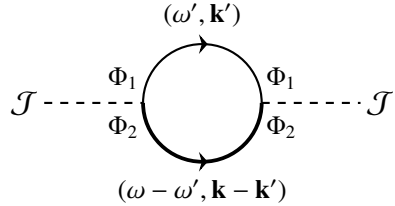
\bes
\begin{tikzpicture}[thick,scale=1.3]
\draw [dashed] (0mm,0) -- (10mm,0);
\draw [mom] (10mm,0) arc (-180:-90:6mm); \draw [mom] (22mm,0) arc (0:-90:6mm); \draw [right] (10mm,0) arc (-180:-90:6mm); 
\draw [right] (10mm,0) arc (180:90:6mm); \draw (22mm,0) arc (0:90:6mm);
\draw [dashed] (22mm,0) -- (32mm,0);
\node at (-2mm,0) {$\mc{J}$}; \node at (34mm,0) {$\mc{J}$}; 
\node at (8mm,2.5mm) {\small {$\Phi_1$}}; \node at (24mm,2.5mm) {\small{$\Phi_1$}};
\node at (8mm,-2.5mm) {\small{$\Phi_2$}}; \node at (24mm,-2.5mm) {\small{$\Phi_2$}};
\node at (16mm,9mm) {\small{($\o', {\bf k}'$)}};
\node at (16mm,-9mm) {\small{($\o-\o', {\bf k}-{\bf k}'$)}};
\end{tikzpicture}
\ees
It is the evaluation of this one-loop integral that results into the non-analytic long-time tails in the $\mc{J} \mc{J}$ two-point function, as we will explore now for the stress tensor correlation functions in our system.

\subsection{Long-time tails in stress tensor correlators}
With the retarded correlation functions known for the two cases $\eta_\parallel \neq 0$ and $\z_1 \neq 0$, it is straight forward to compute the long-time tails in stress tensor two-point functions using the general methodology outlined above. Consider first the case of $\eta_\parallel \neq 0$. As a consequence of the nonlinear interaction terms in eq.\ \eqref{stressnl}, we get the following contribution to the symmetric two-point function of the $T^{xz}$ component of the stress tensor, eq.\ \eqref{fourpoint1},
\be
\label{gint1}
{\rm G}^{\rm int}_{T^{xz} T^{xz}}(t, {\bf x}) = \k^2 \langle \pi_x(t, {\bf x}) \pi_z(t, {\bf x})  \pi_x(0) \pi_z(0)\rangle,
\ee
where 
\be
\k = \f{1}{w_0^2} \left(w_0- M_{\O,\m} B_0^2\right).
\ee
The assumption of Gaussianity for the fluctuations then leads to 
\be
\label{gint2}
{\rm G}^{\rm int}_{T^{xz} T^{xz}}(\o, {\bf k}) 
= \k^2 \int \f{d\o'}{2\pi} \f{d^3k'}{(2\pi)^3} \,\mb{G}^{\rm Sym}_{\pi_x \pi_x}(\o', {\bf k}') \,\mb{G}^{\rm Sym}_{\pi_z \pi_z}(\o-\o', {\bf k} - {\bf k}').
\ee
Putting in the symmetric correlators in eq.\ \eqref{gint2} that follow from the retarded two-point functions eqs.\ \eqref{etaxx} and \eqref{etazz} by the use of the fluctuation-dissipation relation eq.\ \eqref{fdt}, and performing the integrals leads to 
\be
\label{gint3}
{\rm G}^{\rm int}_{T^{xz} T^{xz}}(\o, {\bf k}) = \f{\k^2 T_0^2 w_0^{7/2}}{4\pi \eta_\parallel^{3/2}}  \left( \f{{\rm Coth}^{-1} \sqrt{2}}{\sqrt{2}} - \f{31}{30}\right) \sqrt{|\o|} + \cdots
\ee
The appearance of the non-analytic term proportional to $\sqrt{|\o|}$ is the long-time tail we were interested in finding out. If we Fourier transform the above expression to real time, then we indeed obtain a long-time power-law tail in the two-point function, proportional to $|t|^{-3/2}$. This indicates that after sufficiently long time the two-point correlation function does not die off exponentially rapidly, as would be the case if there were no interactions present, but rather falls off as a power-law.

Similarly, for the case when only $\z_1 \neq 0$, the same stress tensor two-point function receives the non-analytic contribution
\be
\label{gint4}
{\rm G}^{\rm int}_{T^{xz} T^{xz}}(\o, {\bf k}) =  \f{\k^2 T_0^2 w_0^{7/2}}{80 \pi \z_1^{3/2}} \sqrt{|\o|} + \cdots
\ee
which is the long-time tail we were after. One can similarly compute the tails in correlation functions of other components of the stress tensor, and verify that the non-analytic  structure present in eqs.\  \eqref{gint3} and \eqref{gint4} persists. Note that the leading corrections in the presence of the magnetic field are proportional to $M_{\O,\m} B_0^2$ for the stress tensor correlators considered above. For other correlators, the leading dependence on the magnetic field will similarly follow from eq.\ \eqref{stressnl}, as well as from the terms in the linearized equations in \ref{full_lin}.

Another important point to stress is that even though we set up our calculations in the thermodynamic frame, the final results for the long-time tails are independent of this choice. It is so because physically measurable quantities such as the stress tensor are frame invariant objects, and therefore their correlation functions are frame invariant too.

The non-analytic dependence on $\o$ computed in eqs.\ \eqref{gint3} and \eqref{gint4}  above is for the symmetric correlation functions of stress tensor components. Following the fluctuation-dissipation relation eq.\ \eqref{fdt}, this implies that the retarded correlation functions of the stress tensor will have a non-analytic dependence which behaves as $|\o|^{3/2}$. It is the retarded correlation functions which capture the causal response of a system to perturbations, and can be computed by varying the hydrodynamic constitutive relations for the conserved currents with respect to the external sources. Now, in a linearized analysis of fluctuations about a given equilibrium state, one would conclude that the retarded correlators of conserved currents in the limit $\bf{k} \rightarrow 0$ will have a leading behaviour proportional to $\o$, coming from one-derivative terms. This would be followed by a subleading term proportional to $\o^2$ that comes from two-derivative terms in the hydrodynamic derivative expansion. However, as we saw above, the presence of nonlinear interaction terms actually contributes a term proportional to $|\o|^{3/2}$ to the retarded correlator, which is more significant than the terms that come from the second order in the hydrodynamic derivative expansion. We are thus lead to include the effects of nonlinear interaction terms before considering higher derivative terms in the derivative expansion.

Note that the one-loop integral present in the frequency-momentum integrals of eq.\ \eqref{genform1} is in general UV divergent. We have regulated this divergence by introducing a UV cutoff $\L$ in our computations. These cutoff dependent terms have been suppressed in eqs.\ \eqref{gint3} and \eqref{gint4}, to focus on the cutoff independent long-time tails in the correlation functions. The cutoff dependence can be absorbed in the renormalization of the associated transport parameters.

\section{Discussion}
\label{discussion}
In this paper, we studied fluctuations in a relativistic fluid with a conserved global $U(1)$ current in the presence of a background magnetic field, and wrote down the most general linearized hydrodynamic equations for the fluctuations in energy density, charge density and momentum density about a state of global thermal equilibrium (see \ref{full_lin}). The background magnetic field renders the setup anisotropic, which results into several new transport parameters. If there were no anisotropy, there would be three dissipative transport parameters at the first-order in the derivative expansion: the shear viscosity $\eta$, the bulk viscosity $\z$, and the conductivity $\s$. However, the anisotropic background leads to seven dissipative transport parameters in place of three. These are the two shear viscosities $\eta_\parallel, \eta_\perp$, four bulk viscosities $\z_1, \z_2, \eta_1, \eta_2$, and two conductivities $\s_\parallel, \s_\perp$. Only three of the four bulk viscosities are independent due to the Onsager relation that provides a linear condition between three of them. In addition, there are three first-order non-dissipative transport parameters: the Hall conductivity $\tilde{\s}$, and the Hall viscosities $\tilde{\eta}_\parallel, \tilde{\eta}_\perp$. 

With the full set of linearized equations in hand, we were able to focus on different subsectors of transport in the system, in each of which one of the transport parameters dominates over the rest. We in particular focused on two cases, when the shear viscosity $\eta_\parallel \neq 0$ and when the bulk viscosity $\z_1 \neq 0$. In these two subsectors, we computed the retarded two-point correlation functions of the hydrodynamic fluctuations under suitable simplifying assumptions. Further, with the retarded correlators in hand, we went on to compute the effects of nonlinear interaction terms in the hydrodynamic constitutive relations on the correlation functions of the stress tensor for the fluid. We found the appearance of non-analytic long-time tails in these correlation functions, which lead to power-law fall off for the stress tensor correlation functions after long times, as opposed to an exponential decay. 

Several comments are in order. To maintain analytical control we focused on transport in sectors where one of the transport parameters is non-zero, and the rest can be set to vanish. For the purpose of applications to QGP and other physical situations such as dense quark matter in magnetic fields, it would be useful to consider transport when several of the transport parameters are non-zero. This would require a sophisticated numerical approach, which we leave out  for a future work. In particular, with applications to QGP in mind, a logical next step is to develop the formalism and compute the long-time non-analytic behaviour when the magnetic field is dynamical i.e. to make use of the full formalism of relativistic magnetohydrodynamics \cite{Hernandez:2017mch}, which can also be done in the dual formulation in terms of generalized global symmetries \cite{Grozdanov:2016tdf}. Another possible future direction would be to study the long-time tails in the presence of a magnetic field about a state where the fluid is undergoing longitudinal boost invariant expansion, known as Bjorken flow \cite{Martinez:2018wia}. Finally, it would also be interesting to understand the non-analytic long-time tails in the presence of a magnetic field from a holographic perspective \cite{CaronHuot:2009iq, Gursoy:2020kjd}. We hope to report on some of these directions in the near future.

\appendix

\section{The full set of linearized equations}
\label{full_lin}
In this appendix, we present the complete set of linearized hydrodynamic equations about the equilibrium state with temperature $T_0$, chemical potential $\mu_0$, energy density $\e_0$, pressure $p_0$ and fluid velocity $u^\m_0 = (1, {\bf 0})$. The magnetic field is assumed to be in the $+Z$-direction with magnitude $B_0$. The equations are written focusing on the genuine transport parameters that appear at first order in the derivative expansion in an out-of-equilibrium state, ignoring the effects from the static susceptibilities such as $\a_{\rm BB}$, $M_\O$ etc.\ as their effects become subleading when the system is dominated by fluctuations. The equations take the following form, with the fluctuations Fourier transformed i.e. $\d\e (t, {\bf x}) = \d\e(t) \, e^{i {\bf k} \cdot {\bf x}}$, and so on.

\begin{align}
&\del_t \delta\e + i {\bf k} \cdot {\boldsymbol{\pi}} = 0\, , \label{feq1}\\
&\del_t \delta n + \a_2 \left(\s_\parallel k_z^2 + \s_\perp (k_x^2 + k_y^2)\right) \delta n + \a_1 \left(\s_\parallel k_z^2 + \s_\perp (k_x^2 + k_y^2)\right) \delta \e  + \f{in_0}{w_0} k_z \pi_z \nonumber \\
&\qquad+ \f{i}{w_0} \left[(n_0 +B_0 \tilde{\s}) k_x - B_0\s_\perp k_y\right] \pi_x +  \f{i}{w_0} \left[(n_0 +B_0 \tilde{\s}) k_y + B_0\s_\perp k_x\right] \pi_y = 0\, , \label{feq2}\\ 
&\del_t \pi_x + \f{1}{w_0} \left[B_0^2 \s_\perp+(\z_1 + \eta_\perp - \textstyle{\f{1}{3}} \eta_1)k_x^2 + \eta_\perp k_y^2 + \eta_\parallel k_z^2 \right]\pi_x \nonumber\\
&\qquad- \f{1}{w_0} \left[B_0(n_0 + B_0 \tilde{\s}) + \tilde{\eta}_\perp (k_x^2 + k_y^2) + \textstyle{\f{1}{2}} \tilde{\eta}_\parallel k_z^2 - (\z_1 - \textstyle{\f{1}{3}} \eta_1) k_x k_y\right] \pi_y \nonumber \\
&\qquad+ \f{1}{w_0} \left[\z_1+\z_2+\eta_\parallel - \textstyle{\f{1}{3}} (\eta_1+\eta_2) \right]k_x k_z \pi_z -\f{\tilde{\eta}_\parallel}{2w_0} k_y k_z \pi_z \nonumber\\
&\qquad+ i\left[\left(\textstyle{\left(\f{\del P}{\del n}\right)_\e} + B_0 \a_2 \tilde{\s}\right) k_x + B_0 \a_2 \s_\perp k_y\right] \delta n + i \left[\textstyle{\left(\f{\del P}{\del \e}\right)_n} k_x + B_0 \a_1 (k_x \tilde{\s} + k_y \s_\perp)\right] \delta\e = 0\, , \label{feq3}
\end{align}
\begin{align}
&\del_t \pi_y + \f{1}{w_0} \left[B_0^2 \s_\perp+(\z_1 + \eta_\perp - \textstyle{\f{1}{3}} \eta_1)k_y^2 + \eta_\perp k_x^2 + \eta_\parallel k_z^2 \right]\pi_y \nonumber\\
&\qquad+ \f{1}{w_0} \left[B_0(n_0 + B_0 \tilde{\s}) + \tilde{\eta}_\perp (k_x^2 + k_y^2) + \textstyle{\f{1}{2}} \tilde{\eta}_\parallel k_z^2 + (\z_1 - \textstyle{\f{1}{3}} \eta_1) k_x k_y\right] \pi_x \nonumber\\
&\qquad+ \f{1}{w_0} \left[\z_1+\z_2+\eta_\parallel - \textstyle{\f{1}{3}} (\eta_1+\eta_2) \right] k_y k_z \pi_z +\f{\tilde{\eta}_\parallel}{2w_0} k_x k_z \pi_z \nonumber \\
&\qquad+ i\left[\left(\textstyle{\left(\f{\del P}{\del n}\right)_\e} + B_0 \a_2 \tilde{\s}\right) k_y - B_0 \a_2 \s_\perp k_x\right] \delta n + i \left[\textstyle{\left(\f{\del P}{\del \e}\right)_n} k_y + B_0 \a_1 (k_y\tilde{\s} - k_x \s_\perp)\right] \delta\e = 0\, ,\label{feq4}\\
&\del_t \pi_z + \f{1}{w_0} \left[(\z_1+\z_2+\textstyle{\f{2}{3}}(\eta_1+\eta_2)) k_z^2 + \eta_\parallel(k_x^2+k_y^2) \right] \pi_z + \f{1}{w_0} \left[(\z_1 +\eta_\parallel + \textstyle{\f{2}{3}} \eta_1 ) k_x + \textstyle{\half} \tilde{\eta}_\parallel k_y\right] k_z \pi_x  \nonumber\\
&\qquad+ \f{1}{w_0} \left[(\z_1 +\eta_\parallel + \textstyle{\f{2}{3}} \eta_1 ) k_y - \textstyle{\half} \tilde{\eta}_\parallel k_x\right] k_z \pi_y + i \textstyle{\left(\f{\del P}{\del n}\right)_\e} k_z \delta n +  i \textstyle{\left(\f{\del P}{\del \e}\right)_n} k_z \delta\e = 0\, , \label{feq5}
\end{align}
with $$\a_1 = \left(\f{\del \m}{\del \e}\right)_n - \f{\m}{T} \left(\f{\del T}{\del \e}\right)_n,$$ and $$\a_2 = \left(\f{\del \m}{\del n}\right)_\e - \f{\m}{T} \left(\f{\del T}{\del n}\right)_\e.$$

It is straightforward to compute the eigenmodes of small oscillations about the chosen equilibrium state, using the linearized equations above. 
One gets five eigenmodes, of which two are gapped, given by
\be
\o = \pm \f{B_0 n_0}{w_0} - \f{i B_0^2}{w_0} \left( \sigma_\perp \pm i \tilde{\s}\right) + \mc{O}({\bf k}^2),
\ee
while the other three are gapless. For ${\bf k} \parallel {\bf B}$, the three gapless modes are two sound waves and one diffusive mode, given by
\begin{subequations}
\begin{align}
&\o = \pm k v_s - \frac{i}{2} \G_{\parallel} {\bf k}^2\, ,\\
&\o = - i D_{\parallel} {\bf k}^2\, ,
\end{align}
\end{subequations}
where the speed of sound is given by 
$$v_s^2 = \left(\f{\del p}{\del \e}\right)_n + \f{n_0}{w_0} \left(\f{\del p}{\del n}\right)_\e$$
and
\bes
\begin{split}
&\G_\parallel = \f{1}{w_0} \left(\z_1 + \z_2 + \f{2}{3} \left(\eta_1 + \eta_2\right)\right) + \f{1}{v_s^2} \left(\f{\del p}{\del n}\right)_\e \left(\a_1 + \f{n_0}{w_0} \, \a_2\right) \s_\parallel\, ,\\
&D_\parallel = \f{1}{v_s^2} \left(\left(\f{\del p}{\del \e}\right)_n \a_2 -  \left(\f{\del p}{\del n}\right)_\e \a_1\right) \s_\parallel\, .
\end{split}
\ees

For ${\bf k} \perp {\bf B}$, the gapless modes consist of one diffusive, one shear and one sub-shear mode, given by
\begin{subequations}
\begin{align}
&\o = - i D_\perp {\bf k}^2\, ,\\
&\o = - i \f{\eta_\parallel}{w_0} {\bf k}^2\, ,\\
&\o = - i \G_\perp ({\bf k}^2)^2\, ,
\end{align}
\end{subequations}
where
\bes
\begin{split}
&D_\perp = \f{w_0 \left(\left(\f{\del p}{\del \e}\right)_n - n_0 \a_1\right) \s_\perp}{B_0^2 \s_\perp^2 + \left(n_0 + B_0 \tilde{\s}\right)^2}\, ,\\
&\G_\perp = \f{\left(\left(\f{\del p}{\del \e}\right)_n \a_2 -  \left(\f{\del p}{\del n}\right)_\e \a_1\right) \eta_\perp}{B_0^2 \left(\left(\f{\del p}{\del \e}\right)_n - n_0 \a_1\right)}.
\end{split}
\ees
In \cite{Hernandez:2017mch}, the authors study the linearized hydrodynamic equations in terms of the fluctuations $\d T, \d \m$ in the temperature and chemical potential. On the other hand, our analysis above works with the fluctuations $\d\e, \d n$ in the energy and charge density. The two approaches are equivalent, and in particular the modes we have found above are in agreement with the ones reported in \cite{Hernandez:2017mch}, after an appropriate transformation from derivatives with respect to $\e, n$  to derivatives with respect to $\m, T$, which are the independent variables in the grand canonical ensemble.

\section{The susceptibility matrix $\boldsymbol{\chi}$}
\label{suscmat}
In the main text, we focus on two subsectors of the full set of linearized equations in \ref{full_lin}: first when only $\eta_\parallel \neq 0$, and second when only $\z_1 \neq 0$. For both the cases we compute the retarded two-point correlation functions of hydrodynamic fluctuations using linear response theory, see \ref{linrep1} and \ref{linrep2}. For this we need to make use of the susceptibility matrix $\boldsymbol{\chi}$ defined in eq.\ \eqref{defsusc}. This can be computed as follows. The complete set of hydrodynamic fluctuations is
\be
\label{hydrovars}
\Phi = \left(\delta n, \delta \e, \pi_x, \pi_y, \pi_z \right).
\ee
Now the partition function in the grand canonical ensemble is given by 
\be
\mc{Z} = {\rm Tr} \exp\left[-\beta (H- \mu N)\right],
\ee
where $H$ is the Hamiltonian and $N$ is the total charge. Under infinitesimal variations $\delta T, \delta \mu$ and $\delta u^i$ in temperature, chemical potential and velocity, we get the new partition function to be
\be
\label{z2}
\begin{split}
\mc{Z}' = {\rm Tr} &\exp\bigg[-\beta (H- \mu N) + \f{\delta T}{T^2} (H- \mu N) + \f{\delta\m}{T} N +\f{1}{T} \, {\bf \delta u}\cdot{\bf P}\bigg]
\end{split}
\ee
For use in linear response theory, the quantity of interest is the perturbation to the Hamiltonian, which for slowly varying sources can be read off from eq.\ \eqref{z2} as
\be
\d H(t) = - \int d^3x \bigg(\f{\delta T(t,{\bf x})}{T} (\e(t,{\bf x})- \mu \, n(t, {\bf x})) + \d\m(t, {\bf x}) \, n(t, {\bf x}) +\d u^i(t, {\bf x})\, \pi_i(t, {\bf x})\bigg).
\ee
Therefore the sources for the hydrodynamic variables eq.\ \eqref{hydrovars} are given by
\be
\label{hydrosorcs}
\p = \left(\d\m - \f{\m}{T} \d T,\,  \f{\d T}{T}, \, \d u_x, \, \d u_y, \, \d u_z\right).
\ee
One can now compute the susceptibility matrix eq.\ \eqref{defsusc} by differentiating the hydrodynamic variables in eq.\ \eqref{hydrovars} with respect to their sources in eq.\ \eqref{hydrosorcs}, keeping in mind that
$$\d\e = \left(\f{\del\e}{\del\m}\right)_T \left(\d\m - \f{\m}{T} \d T\right) + \left[\m \left(\f{\del\e}{\del\m}\right)_T + T \left(\f{\del\e}{\del T}\right)_\m\right] \f{\d T}{T},$$
$$\d n = \left(\f{\del n}{\del\m}\right)_T \left(\d\m - \f{\m}{T} \d T\right) + \left[\m \left(\f{\del n}{\del\m}\right)_T + T \left(\f{\del n}{\del T}\right)_\m\right] \f{\d T}{T},$$
and $\pi _i = w_0 \, \d u_i$. This gives the susceptibility matrix as
\bes
\boldsymbol{\chi} = \begin{pmatrix}
\left(\f{\del n}{\del\m}\right)_T & \m \left(\f{\del n}{\del\m}\right)_T + T \left(\f{\del n}{\del T}\right)_\m & 0 & 0 & 0\\
\left(\f{\del \e}{\del\m}\right)_T & \m \left(\f{\del \e}{\del\m}\right)_T + T \left(\f{\del \e}{\del T}\right)_\m & 0 & 0 & 0\\
0 & 0 & w_0 & 0 & 0\\
0 & 0 & 0 & w_0 & 0\\
0 & 0 & 0 & 0 & w_0
\end{pmatrix}.
\ees

For computing the long-time tails, we further specialize to an equilibrium state with $\mu_0 = 0$ i.e. the equilibrium charge density vanishes $n_0 = 0$. For systems that respect charge conjugation invariance, this further leads to the vanishing of thermodynamic derivatives such as $(\del n/\del T)_\m = 0, (\del \e/ \del \m)_T = 0$, since they are proportional to $\langle H N \rangle_{\rm conn}$, which vanishes by charge conjugation invariance. Further $\del \e/\del T \equiv c_v = w_0/T v_s^2$, further simplifying the susceptibility matrix above, which can then be used to compute the retarded correlation functions  via eq.\ \eqref{techno} in linear response theory.

\section*{Acknowledgments}
The author would like to thank Pavel Kovtun and Juan Hernandez for very helpful discussions, and for their comments on a draft version of the paper. Research at Perimeter Institute is supported in part by the Government of Canada through the Department of Innovation, Science and Economic Development Canada and by the Province of Ontario through the Ministry of Colleges and Universities.

\bibliography{references}

\begin{thebibliography}{10}
\expandafter\ifx\csname url\endcsname\relax
  \def\url#1{\texttt{#1}}\fi
\expandafter\ifx\csname urlprefix\endcsname\relax\def\urlprefix{URL }\fi
\expandafter\ifx\csname href\endcsname\relax
  \def\href#1#2{#2} \def\path#1{#1}\fi

\bibitem{Kovtun:2012rj}
P.~Kovtun, {Lectures on hydrodynamic fluctuations in relativistic theories}, J.
  Phys. A45 (2012) 473001.
\newblock \href {http://arxiv.org/abs/1205.5040} {\path{arXiv:1205.5040}},
  \href {https://doi.org/10.1088/1751-8113/45/47/473001}
  {\path{doi:10.1088/1751-8113/45/47/473001}}.

\bibitem{Jeon:2015dfa}
S.~Jeon, U.~Heinz, {Introduction to Hydrodynamics}, Int. J. Mod. Phys. E
  24~(10) (2015) 1530010.
\newblock \href {http://arxiv.org/abs/1503.03931} {\path{arXiv:1503.03931}},
  \href {https://doi.org/10.1142/S0218301315300106}
  {\path{doi:10.1142/S0218301315300106}}.

\bibitem{Romatschke:2017ejr}
P.~Romatschke, U.~Romatschke, {Relativistic Fluid Dynamics In and Out of
  Equilibrium}, Cambridge Monographs on Mathematical Physics, Cambridge
  University Press, 2019.
\newblock \href {http://arxiv.org/abs/1712.05815} {\path{arXiv:1712.05815}},
  \href {https://doi.org/10.1017/9781108651998}
  {\path{doi:10.1017/9781108651998}}.

\bibitem{POMEAU197563}
Y.~Pomeau, P.~Résibois,
  \href{http://www.sciencedirect.com/science/article/pii/0370157375900198}{Time
  dependent correlation functions and mode-mode coupling theories}, Physics
  Reports 19~(2) (1975) 63 -- 139.
\newblock \href {https://doi.org/https://doi.org/10.1016/0370-1573(75)90019-8}
  {\path{doi:https://doi.org/10.1016/0370-1573(75)90019-8}}.
\newline\urlprefix\url{http://www.sciencedirect.com/science/article/pii/0370157375900198}

\bibitem{Arnold:1997gh}
P.~B. Arnold, L.~G. Yaffe, {Effective theories for real time correlations in
  hot plasmas}, Phys. Rev. D 57 (1998) 1178--1192.
\newblock \href {http://arxiv.org/abs/hep-ph/9709449}
  {\path{arXiv:hep-ph/9709449}}, \href
  {https://doi.org/10.1103/PhysRevD.57.1178}
  {\path{doi:10.1103/PhysRevD.57.1178}}.

\bibitem{Landau1957}
L.~D. Landau, E.~M. Lifshitz, {Hydrodynamic fluctuations}, JETP 5~(3) (1957)
  512.

\bibitem{PhysRevA.8.423}
P.~C. Martin, E.~D. Siggia, H.~A. Rose,
  \href{https://link.aps.org/doi/10.1103/PhysRevA.8.423}{Statistical dynamics
  of classical systems}, Phys. Rev. A 8 (1973) 423--437.
\newblock \href {https://doi.org/10.1103/PhysRevA.8.423}
  {\path{doi:10.1103/PhysRevA.8.423}}.
\newline\urlprefix\url{https://link.aps.org/doi/10.1103/PhysRevA.8.423}

\bibitem{RevModPhys.49.435}
P.~C. Hohenberg, B.~I. Halperin,
  \href{https://link.aps.org/doi/10.1103/RevModPhys.49.435}{Theory of dynamic
  critical phenomena}, Rev. Mod. Phys. 49 (1977) 435--479.
\newblock \href {https://doi.org/10.1103/RevModPhys.49.435}
  {\path{doi:10.1103/RevModPhys.49.435}}.
\newline\urlprefix\url{https://link.aps.org/doi/10.1103/RevModPhys.49.435}

\bibitem{Crossley:2015evo}
M.~Crossley, P.~Glorioso, H.~Liu, {Effective field theory of dissipative
  fluids}, JHEP 09 (2017) 095.
\newblock \href {http://arxiv.org/abs/1511.03646} {\path{arXiv:1511.03646}},
  \href {https://doi.org/10.1007/JHEP09(2017)095}
  {\path{doi:10.1007/JHEP09(2017)095}}.

\bibitem{Haehl:2015uoc}
F.~M. Haehl, R.~Loganayagam, M.~Rangamani, {Topological sigma models and
  dissipative hydrodynamics}, JHEP 04 (2016) 039.
\newblock \href {http://arxiv.org/abs/1511.07809} {\path{arXiv:1511.07809}},
  \href {https://doi.org/10.1007/JHEP04(2016)039}
  {\path{doi:10.1007/JHEP04(2016)039}}.

\bibitem{Jensen:2017kzi}
K.~Jensen, N.~Pinzani-Fokeeva, A.~Yarom, {Dissipative hydrodynamics in
  superspace}, JHEP 09 (2018) 127.
\newblock \href {http://arxiv.org/abs/1701.07436} {\path{arXiv:1701.07436}},
  \href {https://doi.org/10.1007/JHEP09(2018)127}
  {\path{doi:10.1007/JHEP09(2018)127}}.

\bibitem{Chen-Lin:2018kfl}
X.~Chen-Lin, L.~V. Delacr\'etaz, S.~A. Hartnoll, {Theory of diffusive
  fluctuations}, Phys. Rev. Lett. 122~(9) (2019) 091602.
\newblock \href {http://arxiv.org/abs/1811.12540} {\path{arXiv:1811.12540}},
  \href {https://doi.org/10.1103/PhysRevLett.122.091602}
  {\path{doi:10.1103/PhysRevLett.122.091602}}.

\bibitem{Jain:2020hcu}
A.~Jain, P.~Kovtun, A.~Ritz, A.~Shukla, {Hydrodynamic effective field theory
  and the analyticity of hydrostatic correlators} (11 2020).
\newblock \href {http://arxiv.org/abs/2011.03691} {\path{arXiv:2011.03691}}.

\bibitem{lttsk}
A.~Jain, K.~Jensen, P.~Kovtun, A.~Ritz, A.~Shukla, {Long-time tails in
  diffusive hydrodynamic effective field theory, \textit{In Preparation}}.

\bibitem{Kharzeev:2007jp}
D.~E. Kharzeev, L.~D. McLerran, H.~J. Warringa, {The Effects of topological
  charge change in heavy ion collisions: 'Event by event P and CP violation'},
  Nucl. Phys. A 803 (2008) 227--253.
\newblock \href {http://arxiv.org/abs/0711.0950} {\path{arXiv:0711.0950}},
  \href {https://doi.org/10.1016/j.nuclphysa.2008.02.298}
  {\path{doi:10.1016/j.nuclphysa.2008.02.298}}.

\bibitem{Skokov:2009qp}
V.~Skokov, A.~Illarionov, V.~Toneev, {Estimate of the magnetic field strength
  in heavy-ion collisions}, Int. J. Mod. Phys. A 24 (2009) 5925--5932.
\newblock \href {http://arxiv.org/abs/0907.1396} {\path{arXiv:0907.1396}},
  \href {https://doi.org/10.1142/S0217751X09047570}
  {\path{doi:10.1142/S0217751X09047570}}.

\bibitem{Bzdak:2011yy}
A.~Bzdak, V.~Skokov, {Event-by-event fluctuations of magnetic and electric
  fields in heavy ion collisions}, Phys. Lett. B 710 (2012) 171--174.
\newblock \href {http://arxiv.org/abs/1111.1949} {\path{arXiv:1111.1949}},
  \href {https://doi.org/10.1016/j.physletb.2012.02.065}
  {\path{doi:10.1016/j.physletb.2012.02.065}}.

\bibitem{Kovtun:2003vj}
P.~Kovtun, L.~G. Yaffe, {Hydrodynamic fluctuations, long time tails, and
  supersymmetry}, Phys. Rev. D 68 (2003) 025007.
\newblock \href {http://arxiv.org/abs/hep-th/0303010}
  {\path{arXiv:hep-th/0303010}}, \href
  {https://doi.org/10.1103/PhysRevD.68.025007}
  {\path{doi:10.1103/PhysRevD.68.025007}}.

\bibitem{Hernandez:2017mch}
J.~Hernandez, P.~Kovtun, {Relativistic magnetohydrodynamics}, JHEP 05 (2017)
  001.
\newblock \href {http://arxiv.org/abs/1703.08757} {\path{arXiv:1703.08757}},
  \href {https://doi.org/10.1007/JHEP05(2017)001}
  {\path{doi:10.1007/JHEP05(2017)001}}.

\bibitem{Banerjee:2012iz}
N.~Banerjee, J.~Bhattacharya, S.~Bhattacharyya, S.~Jain, S.~Minwalla,
  T.~Sharma, {Constraints on Fluid Dynamics from Equilibrium Partition
  Functions}, JHEP 09 (2012) 046.
\newblock \href {http://arxiv.org/abs/1203.3544} {\path{arXiv:1203.3544}},
  \href {https://doi.org/10.1007/JHEP09(2012)046}
  {\path{doi:10.1007/JHEP09(2012)046}}.

\bibitem{Jensen:2012jh}
K.~Jensen, M.~Kaminski, P.~Kovtun, R.~Meyer, A.~Ritz, A.~Yarom, {Towards
  hydrodynamics without an entropy current}, Phys. Rev. Lett. 109 (2012)
  101601.
\newblock \href {http://arxiv.org/abs/1203.3556} {\path{arXiv:1203.3556}},
  \href {https://doi.org/10.1103/PhysRevLett.109.101601}
  {\path{doi:10.1103/PhysRevLett.109.101601}}.

\bibitem{Jensen:2013kka}
K.~Jensen, R.~Loganayagam, A.~Yarom, {Anomaly inflow and thermal equilibrium},
  JHEP 05 (2014) 134.
\newblock \href {http://arxiv.org/abs/1310.7024} {\path{arXiv:1310.7024}},
  \href {https://doi.org/10.1007/JHEP05(2014)134}
  {\path{doi:10.1007/JHEP05(2014)134}}.

\bibitem{Kovtun:2019wjz}
P.~Kovtun, A.~Shukla, {Einstein\textquoteright{}s equations in matter}, Phys.
  Rev. D 101~(10) (2020) 104051.
\newblock \href {http://arxiv.org/abs/1907.04976} {\path{arXiv:1907.04976}},
  \href {https://doi.org/10.1103/PhysRevD.101.104051}
  {\path{doi:10.1103/PhysRevD.101.104051}}.

\bibitem{Kovtun:2016lfw}
P.~Kovtun, {Thermodynamics of polarized relativistic matter}, JHEP 07 (2016)
  028.
\newblock \href {http://arxiv.org/abs/1606.01226} {\path{arXiv:1606.01226}},
  \href {https://doi.org/10.1007/JHEP07(2016)028}
  {\path{doi:10.1007/JHEP07(2016)028}}.

\bibitem{Kovtun:2018dvd}
P.~Kovtun, A.~Shukla, {Kubo formulas for thermodynamic transport coefficients},
  JHEP 10 (2018) 007.
\newblock \href {http://arxiv.org/abs/1806.05774} {\path{arXiv:1806.05774}},
  \href {https://doi.org/10.1007/JHEP10(2018)007}
  {\path{doi:10.1007/JHEP10(2018)007}}.

\bibitem{Shukla:2019shf}
A.~Shukla, {Equilibrium thermodynamic susceptibilities for a dense degenerate
  Dirac field}, Phys. Rev. D 100~(9) (2019) 096010.
\newblock \href {http://arxiv.org/abs/1906.02334} {\path{arXiv:1906.02334}},
  \href {https://doi.org/10.1103/PhysRevD.100.096010}
  {\path{doi:10.1103/PhysRevD.100.096010}}.

\bibitem{Abbasi:2015saa}
N.~Abbasi, A.~Davody, K.~Hejazi, Z.~Rezaei, {Hydrodynamic Waves in an Anomalous
  Charged Fluid}, Phys. Lett. B 762 (2016) 23--32.
\newblock \href {http://arxiv.org/abs/1509.08878} {\path{arXiv:1509.08878}},
  \href {https://doi.org/10.1016/j.physletb.2016.09.002}
  {\path{doi:10.1016/j.physletb.2016.09.002}}.

\bibitem{Abbasi:2016rds}
N.~Abbasi, D.~Allahbakhshi, A.~Davody, S.~F. Taghavi, {Hydrodynamic excitations
  in hot QCD plasma}, Phys. Rev. D 96~(12) (2017) 126002.
\newblock \href {http://arxiv.org/abs/1612.08614} {\path{arXiv:1612.08614}},
  \href {https://doi.org/10.1103/PhysRevD.96.126002}
  {\path{doi:10.1103/PhysRevD.96.126002}}.

\bibitem{Ammon:2020rvg}
M.~Ammon, S.~Grieninger, J.~Hernandez, M.~Kaminski, R.~Koirala, J.~Leiber,
  J.~Wu, {Chiral hydrodynamics in strong magnetic fields} (12 2020).
\newblock \href {http://arxiv.org/abs/2012.09183} {\path{arXiv:2012.09183}}.

\bibitem{Hartnoll:2007ih}
S.~A. Hartnoll, P.~K. Kovtun, M.~Muller, S.~Sachdev, {Theory of the Nernst
  effect near quantum phase transitions in condensed matter, and in dyonic
  black holes}, Phys. Rev. B 76 (2007) 144502.
\newblock \href {http://arxiv.org/abs/0706.3215} {\path{arXiv:0706.3215}},
  \href {https://doi.org/10.1103/PhysRevB.76.144502}
  {\path{doi:10.1103/PhysRevB.76.144502}}.

\bibitem{Baggioli:2020edn}
M.~Baggioli, S.~Grieninger, L.~Li, {Magnetophonons \& type-B Goldstones from
  Hydrodynamics to Holography}, JHEP 09 (2020) 037.
\newblock \href {http://arxiv.org/abs/2005.01725} {\path{arXiv:2005.01725}},
  \href {https://doi.org/10.1007/JHEP09(2020)037}
  {\path{doi:10.1007/JHEP09(2020)037}}.

\bibitem{Amoretti:2020mkp}
A.~Amoretti, D.~K. Brattan, N.~Magnoli, M.~Scanavino, {Magneto-thermal
  transport implies an incoherent Hall conductivity}, JHEP 08 (2020) 097.
\newblock \href {http://arxiv.org/abs/2005.09662} {\path{arXiv:2005.09662}},
  \href {https://doi.org/10.1007/JHEP08(2020)097}
  {\path{doi:10.1007/JHEP08(2020)097}}.

\bibitem{Amoretti:2021fch}
A.~Amoretti, D.~Arean, D.~K. Brattan, N.~Magnoli, {Hydrodynamic
  magneto-transport in charge density wave states} (1 2021).
\newblock \href {http://arxiv.org/abs/2101.05343} {\path{arXiv:2101.05343}}.

\bibitem{Huang:2011dc}
X.-G. Huang, A.~Sedrakian, D.~H. Rischke, {Kubo formulae for relativistic
  fluids in strong magnetic fields}, Annals Phys. 326 (2011) 3075--3094.
\newblock \href {http://arxiv.org/abs/1108.0602} {\path{arXiv:1108.0602}},
  \href {https://doi.org/10.1016/j.aop.2011.08.001}
  {\path{doi:10.1016/j.aop.2011.08.001}}.

\bibitem{Grozdanov:2016tdf}
S.~Grozdanov, D.~M. Hofman, N.~Iqbal, {Generalized global symmetries and
  dissipative magnetohydrodynamics}, Phys. Rev. D 95~(9) (2017) 096003.
\newblock \href {http://arxiv.org/abs/1610.07392} {\path{arXiv:1610.07392}},
  \href {https://doi.org/10.1103/PhysRevD.95.096003}
  {\path{doi:10.1103/PhysRevD.95.096003}}.

\bibitem{Martinez:2018wia}
M.~Martinez, T.~Sch\"afer, {Stochastic hydrodynamics and long time tails of an
  expanding conformal charged fluid}, Phys. Rev. C 99~(5) (2019) 054902.
\newblock \href {http://arxiv.org/abs/1812.05279} {\path{arXiv:1812.05279}},
  \href {https://doi.org/10.1103/PhysRevC.99.054902}
  {\path{doi:10.1103/PhysRevC.99.054902}}.

\bibitem{CaronHuot:2009iq}
S.~Caron-Huot, O.~Saremi, {Hydrodynamic Long-Time tails From Anti de Sitter
  Space}, JHEP 11 (2010) 013.
\newblock \href {http://arxiv.org/abs/0909.4525} {\path{arXiv:0909.4525}},
  \href {https://doi.org/10.1007/JHEP11(2010)013}
  {\path{doi:10.1007/JHEP11(2010)013}}.

\bibitem{Gursoy:2020kjd}
U.~G\"ursoy, M.~J\"arvinen, G.~Nijs, J.~F. Pedraza, {On the interplay between
  magnetic field and anisotropy in holographic QCD} (11 2020).
\newblock \href {http://arxiv.org/abs/2011.09474} {\path{arXiv:2011.09474}}.

\end{thebibliography}

\end{document}